\documentclass[a4paper]{article}
\usepackage{float,booktabs}
\usepackage{tablefootnote}
\usepackage{subcaption}
\usepackage{threeparttable}
\usepackage{hyperref}
\newtheorem{definition}{Definition}
\usepackage{INTERSPEECH2021}

\title{Radically Old Way of Computing Spectra: Applications in End-to-End ASR}
\name{Samik Sadhu$^1$, Hynek Hermansky$^{1,2}$}
\address{
  $^1$Center for Language and Speech Processing, Johns Hopkins University, USA\\
  $^2$Human Language Technology Center of Excellence, Johns Hopkins University, USA}
\email{samiksadhu@jhu.edu, hynek@jhu.edu}

\begin{document}

\maketitle
\begin{abstract}
   We propose a technique to compute spectrograms using Frequency Domain Linear Prediction (FDLP) that uses all-pole models to fit the squared Hilbert envelope of speech in different frequency sub-bands.  The spectrogram of a complete speech utterance is computed by overlap-add of contiguous all-pole model responses. A long context window of 1.5 seconds allows us to capture the low frequency temporal modulations of speech in the spectrogram. For an end-to-end automatic speech recognition task, the FDLP spectrogram performs on par with the standard mel spectrogram features for clean read speech training and test data. For more realistic speech data with train-test domain mismatches or reverberations, FDLP spectrogram shows up to 25\% and 22\% relative WER improvements over mel spectrogram respectively.
\end{abstract}\

\section{Introduction}

Since the advent of digital signal processing in speech, speech spectra for automatic speech recognition (ASR) are computed through short time analysis of speech from 10-20 msec speech segments. The spectral dynamics are provided through dynamic features or by concatenation of short-time spectral vectors over appropriate time spans. However, the original spectrograms \cite{koenig1946sound} were derived from energies at outputs of band-pass filters covering the speech spectrum. We return to this original concept of spectrogram and use long term temporal analysis \cite{hermansky1999temporal,athineos2004lp,thomas2008recognition,sadhu2019m,sadhu2019modulation} to directly derive temporal modulations over long segments of speech. \\

Amongst the latter models, Frequency Domain Linear Prediction (FDLP) \cite{herre1996enhancing,athineos2004lp, shenoy2014frequency} is a technique to fit all-pole models to the squared Hilbert envelope of speech with varied degrees of approximation given by the model order leaving behind a frequency modulated component as residual \cite{ganapathy2009applications,kumaresan1999model}.  Firstly, the FDLP model shows similar ``peak-hugging" characteristics like its more well-known dual, Time Domain Linear Prediction (TDLP) \cite{makhoul1975linear} and prioritizes high energy regions of the Hilbert envelope. Secondly, the all-pole approximation of the Hilbert envelope provides a straight-forward way to compute the rate of change of energy with time or \textit{modulation spectrum} of speech. This can be done recursively from the autoregressive coefficients of the all-pole model \cite{oppenheim2004frequency} and allows for selective alleviation of some modulations from the model response when computing the FDLP spectrogram. \\

In section \ref{sec:fdlp_spec} we describe our speech processing technique to obtain the FDLP spectrogram. Subsequently, section \ref{sec:results} analyzes the results from end-to-end ASR models trained with FDLP spectrogram and compares them with the traditional mel spectrogram features.

\section{FDLP spectrogram}
\label{sec:fdlp_spec}

\subsection{Frequency Domain Linear Prediction (FDLP)}
 Given samples of a signal $x$, the squared Hilbert envelope $H$ of $x$ is computed as the squared magnitude of the discrete time analytical signal of $x$ \cite{marple1999computing}. It has been shown that linear prediction analysis of the Discrete Cosine Transform (DCT) of $x$ yields an all-pole model which approximates  the squared Hilbert envelope $H$ with a degree of approximation given by the model order $p$ \cite{ganapathy2012signal}. \\

In linear prediction analysis \cite{makhoul1975linear}, Levinson-Durbin recursion can be used to obtain the model coefficients $\alpha_m,{m=1,2,\dots, p}$ for any specified model order $p$. We define the \textit{FDLP response} $F$ as the Fourier transform of the inverse of this model. Figure 1 shows how the FDLP response fits the energy of the signal $x$.
\begin{figure}[H]
    \centering
    \includegraphics[scale=0.2]{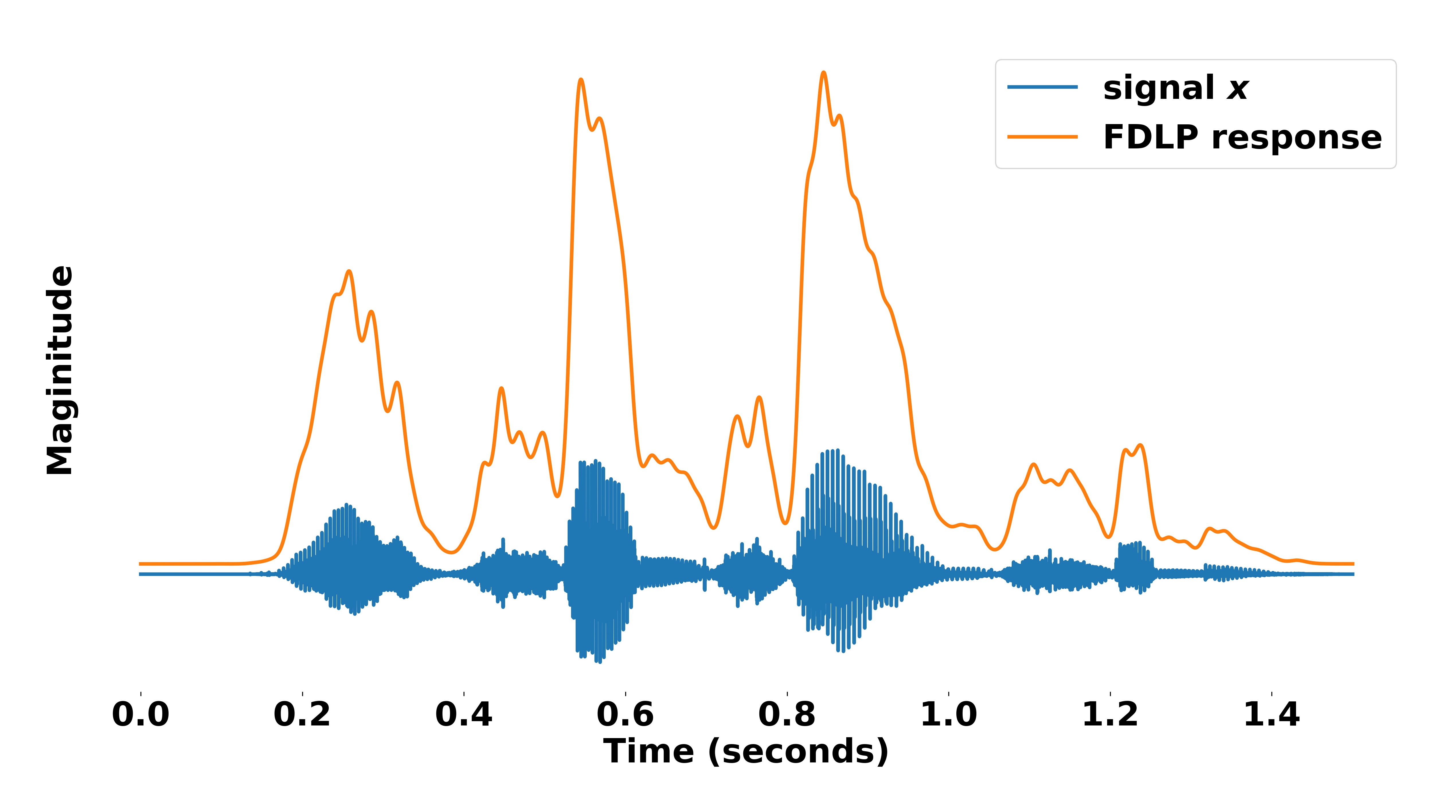}
    \caption{FDLP response of a sample speech signal using an all-pole model of order 150}
    \label{fig:hilbert_envelope}
\end{figure}

\begin{figure*}[tbh]
    \centering
    \hspace{-5pt}
    \includegraphics[trim={2cm 0cm 1cm 0cm},clip,width=13.5cm,height=8.5cm]{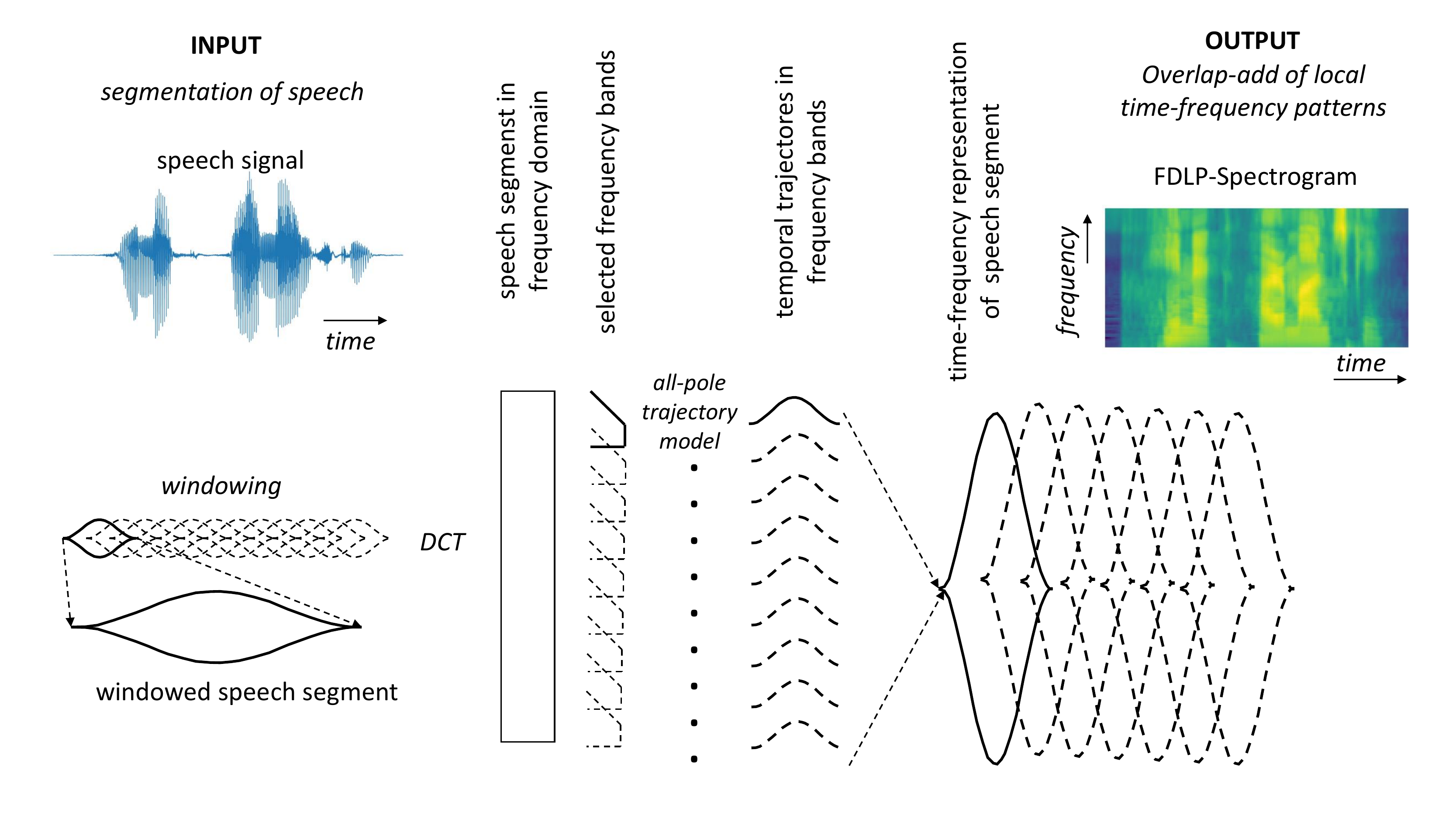}
    \caption{Speech signal is divided into uniform-length shorter speech segments. Each segment is weighted by von Hann window. The windowed segment is transformed to frequency domain through discrete cosine transform (DCT). Auditory-like weights are applied to appropriate parts of the DCT-transformed signal to provide inputs to all-pole modeling. The all-pole models approximate squared Hilbert envelopes of speech segments at different frequencies to yield the local time-frequency pattern of windowed speech segments. Overlap-add technique then yields the total time-frequency pattern of the input speech signal.}
    \label{fig:feature_computation_main}
\end{figure*}
\subsection{Critical-band spectral trajectories using FDLP}
Frequency bands can be formed by windowing of the frequency-domain DCT projection of $x$ as described in the Section \ref{spectrogram}.

\subsection{FDLP to modulation spectrum}
The modulation spectrum captures the variations of the logarithmic energy of the signal $x$ with time \cite{ding2017temporal}. Given that the FDLP response approximates the squared Hilbert envelope $H$, which in turn represents the energy profile of $x$ as a function of time, one reasonable definition of the modulation spectrum of $x$ which we adopt in this work would be as given below.

\begin{definition}
The modulation spectrum of the FDLP response is given by  $\mathcal{M}=IDTFT(\log F)$, where $IDTFT$ is the inverse Discrete Time Fourier Transform.
\end{definition}

Since linear prediction is guaranteed to give stable minimum phase systems, the modulation spectrum can also be computed by recursion directly from the all-pole model coefficients \cite{oppenheim1968homomorphic}  as in eq. \ref{eq:rec}
\begin{equation}
\label{eq:rec}
    \mathcal{M}[m]=\begin{cases}
    0 &\text{ for }  m<0 \\
    \log G &\text{ for } m=0 \\
    \alpha_m + \sum_{i=1}^{m-1}\frac{i}{m} \alpha_{m-i}\mathcal{M}[i] &\text{ for } m>0
    \end{cases}
\end{equation}
where $\alpha_m=0$ for $m>p$. \\
For a $T$ seconds long signal, $\mathcal{M}$ is obtained as projections on cosines that are $\frac{1}{2T}$ Hz apart. Hence modulations upto $F_m$ Hz are captured by the first $2F_mT$ coefficients in $\mathcal{M}$. 

\subsection{Modifying the FDLP response}
Modulation frequencies relevant to speech recognition \cite{hermansky1997modulation} are selected by applying weights $\gamma[m], m=1,2,\dots 2F_mT$ on different cosine projections. $\gamma$ is similar to \textit{cepstral liftering} of TDLP model since the FDLP-derived modulation spectrum is dual to the cepstrum of the TDLP all-pole model. The weighted modulations can be changed to a modified FDLP response as 
\begin{equation}
    \hat{F}=\exp(DTFT(\mathcal{M} \odot \gamma))
\end{equation}
where $\odot$ is the point-wise multiplication operator and DTFT is the Discrete Time Fourier Transform. 

\subsection{Windowing and overlap-add}
The FDLP responses are computed over a fixed time duration of $T$ seconds. However, speech utterances can be of variable duration. The $T$ seconds segments of speech signal are weighted by cosine (von Haan) windows. This allows for applying quarter-window-length Overlap-Add (OLA) \cite{allen1977short} of the FDLP response to concatenate the FDLP responses from the individual speech segment. This operation yields continuous temporal trajectory estimates of the whole speech utterance. The window also de-emphasizes less reliable end-points of the FDLP all-pole approximation. 

\subsection{FDLP spectrogram}
\label{spectrogram}

The FDLP spectrogram is computed from FDLP responses in different frequency sub-bands for a given speech signal $x$. We use 80 cochlear filters \cite{hermansky1990perceptual} equally spaced in the bark scale to separate $DCT(x)$ into frequency sub-bands by point-wise multiplication. To capture low frequency temporal modulations in speech, we use long 1.5 second windows of speech for all-pole model estimation. Assuming a 100 Hz frame rate requirement for the ASR task, the FDLP spectrogram is computed as follows (see figure \ref{fig:feature_computation_main})\\

\begin{enumerate}
    \item Window $x$ using $T=1.5$ seconds long von Haan windows with 25\% overlap
    \item For each windowed signal $x_w$, compute $D_w=DCT(x_w)$
    \item Point-wise multiply $D_w$ with 80 cochlear filter weights to obtain $D_w^{(1)},D_w^{(2)},\dots D_w^{(80)}$.
    \item Do linear predictive analysis of $D_w^{(1)},D_w^{(2)},\dots D_w^{(80)}$.
    \item Compute modulation spectrum $\mathcal{M}_w^{(1)},\mathcal{M}_w^{(2)},\dots \mathcal{M}_w^{(80)}$ from each of the 80 set of linear prediction coefficients using the recursive formulation.
    \item Apply weights $\gamma$ on each modulation spectrum 
    \item Compute log FDLP responses from the weighted modulation spectrum down-sampled to the appropriate frame-rate of 100 Hz. 
    \item The spectrogram for the windowed speech $x_w$ is obtained by forming a $80 \times 100T$ dimensional matrix of the FDLP responses. 
    \item The spectrogram of the complete signal $x$ is computed by OLA of the spectrograms of the time-shifted windows.
\end{enumerate}
The FDLP spectrogram looks similar to mel spectrogram even through the two spectrograms are computed by dual spectro-temporal processing techniques. However, the FDLP spectrogram \textbf{a)} focuses on capturing only energy concentrates in the Hilbert envelope and \textbf{b)} has the added flexibility of choosing different levels of robustness using the all-pole model order and \textbf{c)} manipulating the modulation spectrum. An implementation of FDLP spectrogram is available at \url{https://github.com/sadhusamik/speech\_recognition\_tools}.

\section{Mel spectrogram}
The baseline mel spectrogram (also referred to as Log Filter Bank Energy) features are obtained by short-time analysis of the signal $x$ with 20 ms Hamming windows and a frame-rate of 100 Hz. We compute the magnitude spectrum for each windowed signal. The log spectral energy in 80 mel-scaled triangular filters applied to the magnitude spectrum generates a 80 dimensional vector every 10 ms. These vectors are concatenated over one speech utterance to obtain the mel spectrogram.
\section{Experimental Setup}

\subsection{FDLP spectrogram configuration}
\subsubsection{Window length}
We use $T=1.5$ seconds long von Hann windows to compute the FDLP response. 

\subsubsection{Model order}
A model order of $p$ allows the all-pole model to fit a maximum of $\left \lfloor{\frac{p}{2}}\right \rfloor$ energy peaks of the Hilbert envelope. In section \ref{sec:results} we show how the ASR performance varies with the model order.

\subsubsection{Liftering}
In this work, we only use binary lifters of the form $\gamma[m;a,b]$, where 
\begin{equation}
    \gamma[m;a,b]=\begin{cases}
    1  & \text{ if  } a\leq m \leq b \\
    0  & \text{ else }
    \end{cases}
\end{equation}

Hence, for a window length $T=1.5$ seconds, a lifter $\gamma[m;0,150]$ completely eliminates any cosine projections above $\frac{150}{2\times 1.5}=50$ Hz. Whereas, to eliminate the DC projection, we can apply a lifter $\gamma[m;1,150]$.

\subsection{End-to-end ASR model}
We use the standard transformer based \texttt{espnet1} end-to-end model recipe in the ESPnet \cite{watanabe2018espnet} speech recognition toolkit which uses a joint attention-CTC \cite{kim2017joint} multi-task learning neural network setup. Experiments are done with 12 layers and 6 layers of encoder and decoder respectively with 2048 hidden nodes. A RNN language model is used along-side the acoustic model as in the standard \texttt{espnet1} recipe. The specificities of the training configuration can be found in \url{https://github.com/sadhusamik/speech_recognition_tools/blob/master/e2e/reverb/conf/train.yaml}
\subsection{Data sets}
To analyze the performance of FDLP spectrogram for clean read speech as well as more practical reverberated speech and domain mismatched conditions, we train different ASR models on the following  data sets

\subsubsection{WSJ: clean read speech}
An ASR model is trained with the whole of \texttt{si\_284} data from Wall Street Journal (WSJ) consisting of 73 hours of labelled clean read speech. The model is tested with the clean test set \texttt{test\_eval92}, as well as two artificially corrupted test sets generated by using 20dB of additive street noise and babble noise on \texttt{test\_eval92} respectively. These two additional test sets are named \texttt{street20} and \texttt{babble20} respectively.

\subsubsection{REVERB: noisy reverberated speech}
The performance of FDLP spectrogram on reverberated speech data is evaluated by training an ASR model with \textit{simulated} 8 channel reverberated data from the REVERB challenge \cite{kinoshita2013reverb}. As in the standard ESPnet recipe, this training data is augmented with clean speech from the \texttt{si\_284} training set of WSJ. We test the model on three test sets consisting of \textit{real} reverberated speech data, namely \textbf{a)} \texttt{real\_1ch}: 1 channel speech data with no pre-processing, \textbf{b)} \texttt{real\_1ch\_wpe}: 1 channel speech data with WPE de-reverberation \cite{nakatani2010speech}, \textbf{c)} \texttt{real\_8ch}: 8 channel speech data with WPE de-reverberation and beamforming (using beamformIt \cite{anguera2007acoustic}). \\

For both the data sets, we do not use additional data augmentation techniques (like speed perturbation \cite{ko2015audio} and SpecAugment \cite{park2019specaugment}) which have been used in recent literature to boost end-to-end performance with ESPnet \cite{guo2020recent}.

\section[Results]{Results}
\label{sec:results}
\subsection{Results on WSJ}
Table \ref{tab:mo_and_u} shows how the ASR performance on the clean test set \texttt{test\_eval92} varies with changing model order and lifter configuration. It can be seen that model orders higher than $150$ does not add any significant gain to the ASR performance. 
Previous studies have observed that modulation frequencies in 1-16 Hz range are the most important for ASR as well has human speech cognition \cite{hermansky1997modulation}. In our experiments, we observed noticeable improvements by including cosine projections till 33 Hz (see table \ref{tab:mo_and_u}). However, addition of further modulations adversely affects the ASR performance. Leaving out the DC component of the modulation spectrum slightly degrades performance on this acoustically well controlled WSJ data.\\ 

\begin{table}[htb]
    \caption{Performance on \texttt{test\_eval92} with \textbf{(a)} modulation range 0-33 Hz and various model orders, \textbf{(b)}  model order 150 and various lifter configurations}
    \begin{minipage}{.5\linewidth}
      \caption*{(a)}
      \centering
        \begin{tabular}{@{}cc@{}}
        \toprule
        model order ($p$) & WER \% \\ \midrule
        80          & 5.7    \\
        100         & 5.3    \\
        \textbf{150}         & \textbf{4.8}    \\
        200         & 4.8    \\ \bottomrule
        \end{tabular}
    \end{minipage}%
    \begin{minipage}{.5\linewidth}
      \centering
        \caption*{(b)}
       \begin{tabular}{@{}cc@{}}
        \toprule
        \begin{tabular}[c]{@{}c@{}}lifter \\  configuration ({$\gamma$}) \end{tabular} & WER \% \\ \midrule
        a=0, b=75                                                      & 5.5    \\
        \textbf{a=0, b=100}                                                       & \textbf{4.8}    \\
        a=0, b=150                                                        & 5.2    \\
        a=0, b=300                                                        & 5.1    \\
        a=0, b=450                                                      & 5.0    \\  \midrule 
        a=1, b=100 & 5.3 \\\bottomrule
        \end{tabular}
    \end{minipage} 
    \label{tab:mo_and_u}
\end{table}
Table \ref{tab:compare_wsj} shows a comparison of published state-of-the-art ESPnet performances on WSJ using the same model architecture, our implementation of mel spectrogram and FDLP spectrogram with $p=150$, and modulations in the range 0-33 Hz. It can be seen that FDLP spectrogram performs at-par with the state-of-the-art ESPnet models using mel spectrogram+pitch features. In addition, FDLP spectrogram shows significantly better performance on the noisy mismatched test sets \textit{street20} and \textit{babble20} with up to 25\% relative WER improvements. \\

The constant energy in each frequency sub-bands, represented by the DC component of the modulation spectrum, can easily get corrupted by slowly changing environmental factors or any variations in the microphone characteristics \cite{hermansky1994rasta}. We explore the effects of removing lower cosine projections from our modulation spectrum under reverberated speech conditions in section \ref{sec:reverb}. However, for clean read speech, removing the DC projection (see table \ref{tab:mo_and_u}(b)) can marginally affect ASR performance. \\

\begin{table}[h]
\caption{Comparison of mel spectrogram and FDLP spectrogram performance on WSJ}
\begin{threeparttable}
\begin{tabular}{@{}lccc@{}}
\toprule
Features                  & \multicolumn{3}{c}{WER \%}                                 \\ \midrule
                          & \scriptsize{\texttt{test\_eval92}} & \scriptsize{\texttt{street20}} & \scriptsize{\texttt{babble20}} \\ \cmidrule(l){2-4} 
Guo et al. \cite{guo2020recent} \tnote{*}       & 4.9                & -                 & -                 \\ \midrule
our mel spectrogram & 5.1                & 24.7              & 75.2              \\
FDLP spectrogram             & 4.8                & 20.4              & 56.1              \\ \bottomrule
\end{tabular}

\begin{tablenotes}\footnotesize
\item[*] This result uses mel spectrogram + pitch as features as well as SpecAugment data augmentation technique.
\end{tablenotes}
\end{threeparttable}
\label{tab:compare_wsj}
\end{table}

\subsection{Results on REVERB}
\label{sec:reverb}
The effect of reverberation is captured in low modulation frequencies. Table \ref{tab:mo_and_u2}(a) shows the performance of the ASR model on the \texttt{real\_8ch} test set trained with FDLP spectrogram using a model order of 150 and different ranges of low cosine projections removed. \\
\begin{table}[h]
    \caption{Performance on \texttt{real\_8ch} with model order 150 as a result of \textbf{(a)} removing low modulations \textbf{(b)} including higher modulations from FDLP spectrogram}
    \begin{minipage}{.5\linewidth}
      \caption*{(a)}
      \centering
               \begin{tabular}{@{}cc@{}}
        \toprule
        \begin{tabular}[c]{@{}c@{}}lifter \\  configuration ({$\gamma$}) \end{tabular} & WER \% \\ \midrule
        a=0, b=100                                                      & 8.5    \\
        \textbf{a=1, b=100}                                                       & \textbf{7.9}   \\
        a=2, b=100                                                        & 8.0    \\ \bottomrule
    \end{tabular}
    \end{minipage}%
    \begin{minipage}{.5\linewidth}
      \centering
        \caption*{(b)}
       \begin{tabular}{@{}cc@{}}
        \toprule
        \begin{tabular}[c]{@{}c@{}}lifter \\  configuration ({$\gamma$}) \end{tabular} & WER \% \\ \midrule
        a=1, b=75                                                      & 8.4    \\
        a=1, b=100                                                       & 7.9    \\
        a=1, b=150                                                       & 7.7    \\
        a=1, b=300                                                       & 7.8    \\
        \textbf{a=1, b=450}                                                        & \textbf{7.2}   \\
        a=1, b=600                                                        & 7.7\\  \bottomrule
    \end{tabular}
    \end{minipage} 
    \label{tab:mo_and_u2}
\end{table}

On the other hand, table \ref{tab:mo_and_u2}(b) shows that including higher cosine projections up to 450 Hz achieves a better performance. The reverberated signal is generated by convolving clean speech with room impulse responses that smooth out sudden transitions that characterize plosive sounds. Addition of these higher modulations better preserves remaining abrupt transitions in the FDLP spectrogram. In fact, in our experiments, plosive characters like B and P show up to 5\% reduction in recognition error when modulations between 100 to 150 Hz are included in the FDLP spectrogram.

\begin{table}[h]
\caption{Comparison of mel spectrogram and FDLP spectrogram performance together with results from recent literature on REVERB data.}
\begin{threeparttable}
\begin{tabular}{@{}lccc@{}}
\toprule
Features                  & \multicolumn{3}{c}{WER \%}                                 \\ \midrule
                          & \scriptsize{\texttt{real\_8ch}} & \scriptsize{\texttt{real\_1ch}} & \scriptsize{\texttt{real\_1ch\_wpe}} \\ \cmidrule(l){2-4} 
Guo et al. \cite{guo2020recent} \tnote{*}    & 14.3                   & -                 & -                 \\ 
Zhang et al. \cite{zhang2020end} \tnote{$\dagger$}    & 10.0                   & -                 & -                 \\ \midrule
our mel spectrogram & 9.2                & 23.2             & 20.7              \\
FDLP spectrogram             & 7.2                & 19.4              & 18.0              \\ \bottomrule
\end{tabular}
\begin{tablenotes}\footnotesize
\item[*] \footnotesize{This result uses mel spectrogram + pitch as features as well as speed perturbation data augmentation technique.}
\item[$\dagger$] \footnotesize{Unifies de-reverberation and ASR under one architecture.}
\end{tablenotes}
\end{threeparttable}
\label{tab:compare_reverb}
\end{table}

Table \ref{tab:compare_reverb} shows that FDLP spectrogram has a 22\% relative WER improvement over our mel spectrogram features. Additionally, ASR performance using FDLP spectrogram \textit{without} WPE front-end de-reverberation is better than  mel spectrogram \textit{with} WPE de-reverberation. Thus, FDLP spectrogram is more effective at dealing with the effects of reverberation as compared to WPE. Using WPE front-end processing as well as FDLP spectrogram features yield a 13\% relative WER reduction over mel spectrogram with WPE de-reverberation.

\section{Conclusions}
In this work we described a way to compute spectrograms using Frequency Domain Linear Prediction with several robustness benefits. The proposed spectrogram shows significant improvements over mel spectrogram for domain mismatched train-test scenarios as well as noisy, reverberated speech data and is better at handling the effects of reverberation compared to WPE alone.

\section{Acknowledgements}
This work was funded by a faculty gift from Google Research and by the second author's research support from the JHU Human Language Technology Center of Excellence.
\\
\\

\bibliographystyle{IEEEtran}

\bibliography{mybib}


\end{document}